%% file: main.tex
\pdfoutput=1

\documentclass[11pt]{article}

\usepackage[final]{acl}

\usepackage{times}
\usepackage{latexsym}

\usepackage[T1]{fontenc}

\usepackage[utf8]{inputenc}

\usepackage{microtype}

\usepackage{inconsolata}

\usepackage{graphicx}

\include{latex/includes/includes}

\usepackage{listings}
\usepackage{xcolor}
\usepackage{tcolorbox}
\usepackage{caption}

\newcommand{\dataset}{\texttt{DiSCo}}
\newcommand{\alignment}{\texttt{LPO}}
\newcommand{\Task}{Secure Code Generation}
\newcommand{\task}{secure code generation}

\title{Teaching an Old LLM Secure Coding: \\
Localized Preference Optimization on Distilled Preferences}

\author{
Mohammad Saqib Hasan$^{1}$ \quad Saikat Chakraborty$^{2}$ \quad Santu Karmaker$^{3}$ \\
\textbf{Niranjan Balasubramanian}$^{1}$\\
$^1$Stony Brook University, $^2$Microsoft Research, $^3$University of Central Florida \\
$^1$\texttt{\{mdshasan,niranjan\}@cs.stonybrook.edu}, 
$^2$\texttt{saikatc@microsoft.com}, 
\\$^3$\texttt{santu@ucf.edu}
}

\date{}

\begin{document}
\maketitle
\input{latex/sections/abstract}
\input{latex/sections/introduction}
\input{latex/sections/related_works}

\input{latex/sections/background}

\input{latex/sections/synthetic_data}
\input{latex/sections/preference_optimization}
\input{latex/sections/experimental_setup}

\input{latex/sections/results_and_analysis}
\input{latex/sections/conclusion}
\input{latex/sections/limitations}
\input{latex/sections/ethical_consideration}
\input{latex/sections/acknowledgements}

\bibliography{latex/custom}
\input{latex/sections/appendix}

\end{document}

%% file: latex/includes/includes.tex
\usepackage{arydshln}
\usepackage[font=small,labelfont=bf,tableposition=top]{caption}
\usepackage{hyperref}
\usepackage{boldline}
\usepackage{color,colortbl}
\usepackage{bigstrut}
\usepackage[ruled]{algorithm2e}
\usepackage{amsmath,amsfonts}
\usepackage{amsmath}
\usepackage{array}
\usepackage{algorithmic}
\usepackage{balance}
\usepackage{blindtext}
\usepackage{booktabs}
\usepackage{caption}
\usepackage{hyperref}
\usepackage[noabbrev]{cleveref}
\usepackage{comment}
\usepackage{enumitem}
\usepackage{eqparbox}
\usepackage{fancybox}
\usepackage{fancyvrb}
\usepackage{framed}
\usepackage{ifthen}
\usepackage{listings}
\usepackage{mathrsfs}
\usepackage{mdwmath}
\usepackage{mdwtab}
\usepackage{multirow}
\usepackage{pifont}
\usepackage{stfloats}
\usepackage{textcomp}
\usepackage{times}
\usepackage{url}
\usepackage{xspace}
\usepackage[normalem]{ulem}
\usepackage[framemethod=TikZ]{mdframed}
\usepackage{caption}
\usepackage{subcaption}
\usepackage{tabularx}
\usepackage[switch]{lineno}

\usepackage{titlesec}

\usepackage{tcolorbox}
\tcbuselibrary{listings,skins}
 
\usepackage{mathtools}
\usepackage{blindtext}
\usepackage{lstautogobble}

\usepackage{latexsym}

\usepackage[T1]{fontenc}

\usepackage[utf8]{inputenc}

\usepackage{microtype}

\usepackage{inconsolata}

\usepackage{graphicx}

\usepackage{svg}

%% file: latex/sections/abstract.tex
\begin{abstract}
LLM generated code often contains security issues. We address two key challenges in improving secure code generation. First, obtaining high quality training data covering a broad set of security issues is critical. To address this, we introduce a method for distilling a preference dataset of insecure and secure code pairs from frontier LLMs, along with a security reasoning that explains the issues and the fix. The key idea here is to make use of security knowledge sources to devise a systematic prompting strategy that ensures broad coverage. Second, aligning models to secure code requires focusing on localized regions of code. Direct preference optimization methods, like SimPO, are not designed to handle these localized differences and turn out to be ineffective. We address this with a new localized preference optimization algorithm that masks the security related tokens in both the winning (secure) and losing (insecure) responses. To prevent loss in code quality, we also add a regularizer. Evaluations show that both training on our dataset, \dataset{}, and the new preference optimization algorithm, \alignment{}, yield substantial reductions in code insecurity while also improving overall code quality. Code and dataset are available at \href{https://github.com/StonyBrookNLP/disco-lpo}{https://github.com/StonyBrookNLP/disco-lpo}.

\end{abstract}


%% file: latex/sections/introduction.tex
\section{Introduction}
\label{sec:introduction}

LLMs are increasingly used for coding due to their advanced programming abilities. GitHub Copilot, a popular coding assistant, had over 1.2 million subscribers in 2021~\cite{copilot_usage}, while one survey interviewed $500$ developers of whom $92\%$ state they use AI for coding\cite{ai_usage_survey}. Training smaller models to be effective coders will further improve the adoption of these advances \cite{distill_code_llms}. However, it is important to make sure AI generated code is secure, i.e., it does not contain insecure behavior identified under Common Weakness Enumeration (CWE) \cite{cwe} classification. 
Multiple studies show that a large percentage of AI generated code ($40$-$76\%$) can be insecure~\cite{chatgpt_code_security, asleep_at_the_keyboard}---highlighting the need for reducing security issues in generated code.


Previous works sought to improve security of LLM generated code by tuning them on aggregated data of secure code and/or insecure code~\cite{sven,safecoder}. 
This, however, presents two fundamental challenges: (i) Accumulating training data at scale covering diverse security issues is difficult, expensive, and requires domain expertise.
As a result, many opt for automatic curation from open-source repositories. However, such data tends to be noisy and have low CWE coverage. Filtering for noise further reduces the overall dataset size.
(ii) Designing appropriate alignment techniques for \task{} is challenging.
Standard fine-tuning paradigms, while useful, are not optimized for learning secure coding, as they give equal importance to security relevant and non-relevant tokens during training. 
Models are also not trained to reason about security when generating code.
Preference optimization methods (e.g. \cite{dpo,simpo}) provide a better formulation for teaching models preference of secure over insecure code. However, unlike standard preference data, the distinction between secure and insecure code is often localized to a small region within the code. We want alignment that exploits this \textit{locality} characteristic of the problem. 

Our solution addresses these challenges through two important contributions.
\paragraph{1) Distilling Secure Code from Frontier LLMs (\dataset{}):} We synthesize training data using frontier LLMs to align smaller LLMs towards \task{}. The difficulty here is that directly prompting frontier models is ineffective in multiple ways. While frontier LLMs generate high-quality code for many coding task prompts, the code they generate is not always secure. Moreover, we want synthetisized data to have high CWE coverage.

Addressing these, we propose a prompting strategy which makes use of existing security knowledge sources. These provide a way to first nudge the LLMs towards generating code with known security issues and then generate a fixed version that removes these issues. In addition, we use static security analyzers to identify any remaining issues and ask the LLMs to refine code based on this feedback. Using this pipeline we create \dataset, a large training set of $10$k preference pairs i.e. insecure and secure code along with a reasoning on the issue and the fix useful for secure code alignment.

\paragraph{2) A novel Localized Preference Optimization (\alignment) algorithm:}
Prior works for \task\ have primarily relied on supervised fine-tuning (SFT) solutions, augmenting them with unlikelihood training to favor secure code over insecure code~\cite{sven,safecoder}. Recent advances in preference optimization algorithms provide a more natural formulation for aligning models to prefer secure code. However, unlike standard preference data in other domains (e.g. summarization), the security issues are often highly localized to small regions in the overall code. In other words, the secure and insecure code differ only in small but significant ways, and are near identical in other parts. Existing preference optimization algorithms such as SimPO, the one we build upon, are not well-suited for optimizing on these type of preferences. 

We introduce a new localized preference optimization (\alignment) that propagates loss to security relevant tokens. However, this strategy by itself can lead to lower code quality. We redress this by adding an SFT loss over other tokens as regularizer. 

Our empirical evaluations demonstrate the utility of \dataset{} for improving \task{}. In addition to covering a broad set of CWEs, training on \dataset{} also requires models to generate security reasoning (i.e. identify potential issues and the possible fixes) before generating code. These lead to significant improvements even for standard SFT. The new \alignment{} algorithm further reduces security issues drastically --from $19-40\%$ reductions on four secure coding benchmarks, while also improving code quality -- from $3-10\%$ improvements on two coding benchmarks.

To summarize, this paper makes the following contributions to advance \task{}: 
(i) data distillation pipeline to generate preference data from frontier LLMs using prompts augmented with security domain info. 
(ii) a new training formulation for LLMs whereby they reason on potential security issues before generating code.
(iii) a novel preference optimization algorithm specifically designed for the localized nature of the preferences in \task{}.
(iv) a large distilled dataset of $10$k instances in Python for secure code generation, covering a wider range of known security issues than previous datasets.

%% file: latex/sections/related_works.tex
\section{Related Works}
Code generation quality has greatly improved in both general purpose frontier LLMs (e.g. GPT-4o, Claude) as well as in smaller LLMs, pretrained on public codebases. For example, code LLMs such as CodeLLama~\cite{codellama}, and StarCoder~\cite{Starcoder,starcoder2}, show impressive performance on tasks like basic coding~\cite{humaneval,mbpp}, competitive programming~\cite{apps} and fixing code repository issues \cite{repocorder,swebench}.
Instruction tuning~\cite{instruction_tuning} on synthesized (instruction,code) corpora distilled from frontier models (e.g. GPT-3.5 and GPT-4) has also yielded consistent improvements in overall code quality~\cite{dolphcoder,wavecoder,octopack,code_really_good_data,selfcodealign,beyond_code,code_comments}.
Our work relates to two extensions of this trend: (i) aligning LLMs to improve its overall quality and other specific aspects, (ii) other methods that improve security of generated code.

\paragraph{Aligning LLMs for Code Generation:}
\citet{deliberative_alignment} enhance this synthesis further by distilling safety and utility reasoning as part of training corpus. Others incorporate reinforcement learning (RLHF) \cite{rlhf} for better quality code generation \cite{coderl,stepcoder}. 
Preference optimization algorithms~\cite{dpo,simpo}, which directly optimize policy models, offer a simpler alternative to RLHF, since they do not need separate memory-intensive reward models. Pivotal Token Search \cite{phi4_technical_report} focuses optimization on relevant tokens by estimating their contribution to probability of the overall target response. \citet{synthetic_evolution} improves preference datasets through critique models and code perturbation. While people have trained models for better code generation, ours is the first work that align models towards \task{} using preference tuning over relevant tokens.

\paragraph{Security of Code Generation:}
To analyse security of LLM generated code, works introduce benchmarks like Security Eval~\cite{securityeval}, LLMSecEval~\cite{llmseceval} and outcome-based security evaluation benchmark CWEval~\cite{cweval}. Recent literature propose code security alignment techniques via contrastive prefix tuning~\cite{sven}, unlikelihood tuning~\cite{safecoder}, LLM refinement~\cite{pythonseceval_security_refine}, retrieval augmented generation~\cite{seccoder}, synthesizing data using verifier feedback~\cite{hexacoder} and creating synthetic datasets using LLM parametric knowledge\cite{prosec}. In contrast, we use a more comprehensive data domain knowledge incorporated distillation pipeline and alignment method that incorporates reasoning over security and preference optimization over security relevant tokens for \task{} task.

%% file: latex/sections/background.tex
\section{Secure Code Generation}

Multiple studies show that LLM generated code has security issues as defined by Common Weakness Enumeration \cite{cwe} classification. For example, \citet{chatgpt_code_security} discovered this insecurity rate was 76\% for ChatGPT with only 43\% fixable by prompting, \citet{asleep_at_the_keyboard} stated that Copilot generates on average 40\% insecure code and \citet{user_study_security} found that users generate 10\% more insecure code when using LLMs. This problem can be solved via refinement but is expensive. Hence, an ideal model should directly generate code that is both functional and secure.

\subsection{Task Definition}
Formally, given a prompt $x$, the \task{} task is to generate code $y$ that maximizes utility while minimizing security issues. Utility is calculated via code compilation and unit tests, and security by using security analysis tools (CodeQL~\cite{codeql}, Bandit~\cite{bandit}, etc.).

\subsection{Challenges in \task}
Secure code generation poses two broad challenges: (i) acquiring useful training data and (ii) effectively aligning models towards \task{}.





\paragraph{Acquiring training data:} 
Training for \task{} requires tuples of (\texttt{prompt}, \texttt{secure code}, \texttt{insecure code}). 
Manually curating such data at scale is expensive, difficult, and requires domain expertise.
Prior work, instead, relied on automatic curation and labeling from publicly available resources, like GitHub, using tools and techniques like matching security keywords in the documentation and/or security analyzers~\cite{cvefixes,bigvul}.This led to poor quality data due to poor documentation and tool inaccuracy~\cite{sven}. They also possess other drawbacks: the gathered dataset is often small, vulnerability coverage is limited, and security issues are specific to codebases and non-generalizable~\cite{safecoder}. 

\paragraph{Aligning LLMs for \task:}
One can directly align LLMs via supervised fine-tuning towards generating secure code. Previous works have improved upon this by making use of insecure code via contrastive/unlikelihood training \cite{sven, safecoder}. Preference optimization algorithms (e.g., DPO~\cite{dpo}, SimPO~\cite{simpo}), which provide a more natural formulation for preference learning, have demonstrated superior alignment performance in other problem domains. However, unlike preference data in other domains, the difference between secure (winning response) and insecure code (losing response) is localized to small regions in the code. This means secure and insecure code are mostly similar but have small but important differences. Standard preference optimization solutions (e.g. SimPO) are not well-suited for this setup as they do not utilize the characteristics of the problem. Furthermore, such alignment can to lead to lower code utility as model might overfit towards code security, i.e., not generating certain 
insecurity prone functions (\texttt{os.popen}) over security concerns despite necessity for the task.

\subsection{Our solution}
We address these challenges by introducing a scalable dataset distillation method and a localized preference optimization algorithm (see below). 

\paragraph{\dataset{}:} We show that we can use frontier LLMs to distill (secure, insecure) code pairs along with the reasoning on insecurities and fixes. Key strengths of our approach are: i) control of synthesis pipeline to ensure broad coverage of a broad class of CWEs, ii) reduced noise in dataset by using security analysis tools, and iii) having security reasoning  highlighting the difference between secure and insecure code usable during finetuning.

\paragraph{\alignment{}:} We introduce a new preference optimization loss accounting for the localized nature of the difference between secure and insecure code. This loss function focuses on the difference between the log probabilities for the security-related tokens in the secure code and insecure code during tuning. Furthermore, to reduce loss in code generation quality, we introduce regularization via supervised fine-tuning over the other tokens.

\begin{figure*}[!htbp]
  \centering
\includegraphics[width=1.0\textwidth]{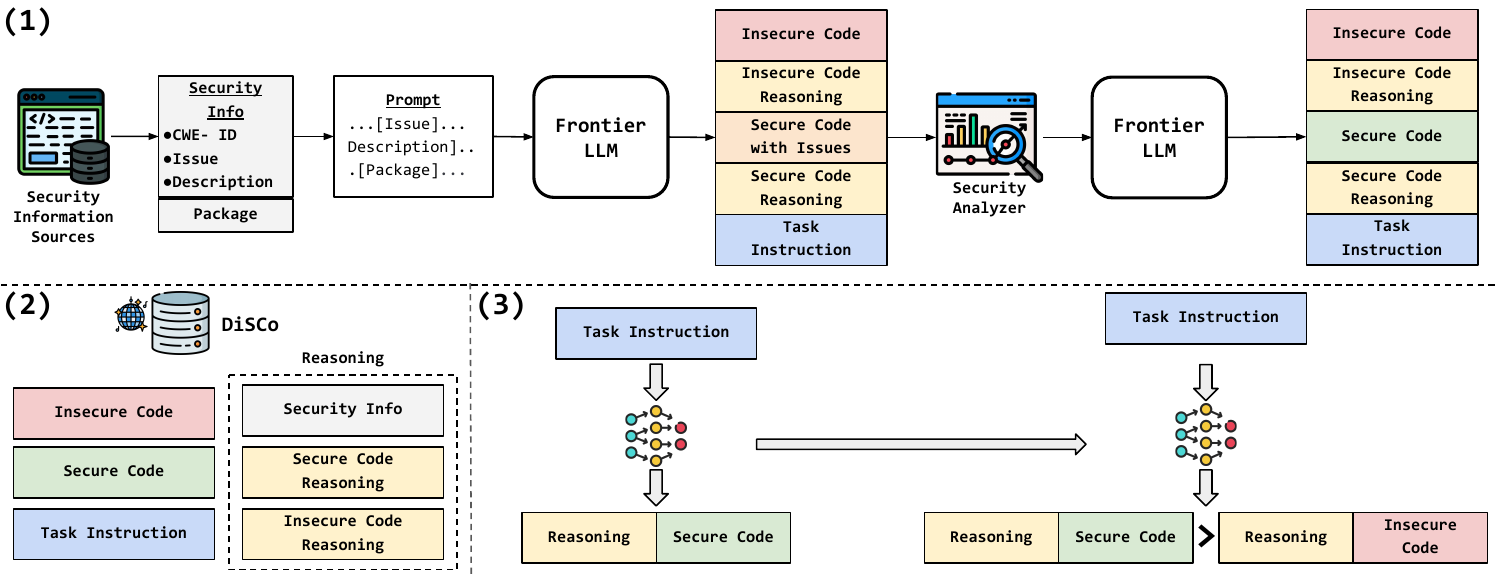}
  \caption{\small{\textbf{Our Methodology}:\texttt{(1)}We extract security information from open domain sources such as \texttt{(CWE-ID, Issue, Description)} and \texttt{Package} info to create prompts. These prompts are fed into frontier models to generate our dataset \dataset{}. To improve security of generated secure code, we feed it through an analyzer, obtain feedback and prompt the LLM to refine it. \texttt{(2)} Our final \dataset{} contains pairs of \texttt{(Task Instruction, Secure Code Insecure Code)}. To create \texttt{Reasoning} element, we combine \texttt{Security Info}, \texttt{Insecure Code Reasoning} and \texttt{Secure Code Reasoning}. \texttt{(3)} To align models towards \task{} using \dataset{}, models are first supervise finetuned to produce \texttt{Reasoning + Secure Code} given \texttt{Task Instruction}. Then, this model is further tuned using our preference tuning loss function \alignment{} that trains model to prefer generating \texttt{(Reasoning + Secure Code)} over \texttt{(Reasoning + Insecure Code)} given the \texttt{Task Instruction}.
  }
  }
  \vspace{-1em}
  \label{Fig:synthetic_data}
\end{figure*}

%% file: latex/sections/synthetic_data.tex
\section{\dataset: Distilling Secure Code Alignment Dataset}
We prompt frontier LLMs to generate a dataset with instances of the form $(x,y^{-},r^{-},y^{+},r^{+})$ where $x$ is the task prompt, $y^{-}$ and $r^{-}$ are insecure code and reasoning on its security, and  $y^{+}$ and $r^{+}$ are the secure code and reason on how it is secure.
Using simple prompts (e.g. [\texttt{Generate an insecure code and corresponding secure code with reasoning.}]) to generate such data is ineffective for two reasons: (i) LLMs often churn out \emph{easy} instances covering the common CWE classes, and (ii) the generated $y^{+}$ may not be secure. We solve these problems by using a security knowledge base during prompting and a \emph{refinement} step using external security analyzers.


\subsection{Generating Prompts for Distillation}
\label{subsec:distillatin-prompting}
For sampling high-quality instances covering various CWEs, we use a security knowledge base curated from information hubs and documents
(CWE website, security analyzer documentation like CodeQL etc.)
In this way, we have a dataset of (\texttt{CWE-ID}, \texttt{Issue}, \texttt{Description}) security info tuples, where \texttt{Issue} and \texttt{Description} are short and long explanations of a code vulnerability and \texttt{CWE-ID} is the CWE category of the security issue. We also extract a list of \texttt{Package} instances (e.g. \texttt{requests}, \texttt{os}) that may contain security issues. These information act as slots that fit in our prompt templates for automatic prompt generation. Through combinations of elements from security info and \texttt{Package} sets, we sample large number of prompts that ensures high CWE and application coverage.



\subsection{Distillation with Refinement:} \label{subsec:distillation-with-refinement} Outputs from frontier LLMs can be erroneous. In our usescase, example of such are security reasoning and code mismatch, insecure and secure code are having different utility, or the prompt misaligned with code. Also, we observed that in $37.4\%$ examples, generated secure code  $y^{+}$ contained security issues either because the actual issue was unfixed, or other issues outside the scope of the prompt exist.
To mitigate this, we add a refinement step. Key idea here is to use security analyzers on the generated code to obtain feedback and then prompt the frontier LLM to reflect and refine its output with this information. We observe that one such refinement step reduces the percentage of data points containing such security issues from $37.4\%$ to $12.7\%$. Multiple refinements ($3$ iterations) can further reduce this to $9.4\%$. However, the overall quality of the resulting synthetic data worsens due to \textit{overengineering} for code security. (See Section \ref{sec:analyses} for more details). Relevant prompts are listed in Appendix \ref{appendix:prompts}.

%% file: latex/sections/preference_optimization.tex
\section{Preference Optimization for \Task}

We use the \dataset{} to train LLMs for \task{} in two stages: (i) supervised fine-tuning to generate reasoning $R$ and secure code $y^{+}$  given prompt $x$ , and (ii) preference tuning to favor $y^{+}$ over $y^{-}$ given $x$. We introduce \emph{Localized Preference Optimization} (\alignment{}), a novel preference optimization algorithm that exploits knowledge of localization of areas that pertain to preference.

\subsection{Supervised Fine-tuning with Security Reasoning}
We first train LLMs to generate secure code $y^{+}$ and security reasoning $R$ given instruction $x$. $R$ is created by concatenating the \texttt{CWE-ID}, \texttt{Issue}, \texttt{Description}, $r^{+}$ and $r^{-}$ in this order (template in Appendix \ref{appendix:prompts}). Requiring the models to generate this security reasoning $R$ nudges them to consider the possible security issues, what insecure code could look like, and then generate the secure code in relation to these considerations. Using this dataset $D_{sft}$ of $(x, R, y^{+})$ tuples, the target LLM $\pi_{\theta}$ is optimized using the following log-likelihood loss:
\begin{align*}
    \mathcal{L}_{SFT} = -\mathbb{E}_{(x,y^{+},R)\sim D} \log \pi_{\theta}(y^{+},R|x)
\end{align*}
\subsection{Localized Preference Optimization}
\label{sec:lpo}
We want models to prefer secure code over insecure code. In our scenario, this means we want models ($\pi_\theta$) that prefer $y^{+}$ over $y^{-}$ over for the same prompt $x$ i.e., we want models with $\pi_{\theta}(y^{+},R|x) > \pi_{\theta}(y^{-},R|x)$. In regular preference optimization like SimPO, loss is measured over all code tokens. However, in our setting only a small handful of tokens determine the presence (or absence) of the security issue. 
Hence, propagating loss across all tokens will dampen the signals from the important security relevant tokens most influential to learning.
To remedy this, we introduce a new preference loss that \emph{localizes} on the security related tokens. We do this by introducing two binary mask vectors $m^{+}$ and $m^{-}$ for $y^{+}$ and $y^{-}$ respectively. Value of $1$ in $m$ denotes security-relevant token and $0$ otherwise.
$m^{+}$ and $m^{-}$ are constructed by identifying the differing tokens between $y^{+}$ and $y^{-}$ by computing the delta between them. The reasoning trace $R$ is same when optimizing for $y^+$ and $y^-$ and therefore masked out, i.e., the value $m$ is $0$ over $R$.

This localization insures loss is propagated only for security related tokens.
However, as we show later, such training causes model to lose code generalization as models tend to hack the reward function \cite{reward_hacking} by generating non-parseable or incoherent code (which can be interpreted as secure code by analyzers).
To adjust for this, we introduce a regularizer, which is the $\mathcal{L}_{sft}$ over the rest of the $y^{+}$ tokens, calculated via complement of the mask $\overline{m}^{+}$. 

Formally, the \alignment{} loss function is given by:
\begin{align*}
    \mathcal{L}_{\texttt{LPO}} &= - \mathbb{E}_{(x,y^{+},y^{-},R)\sim D}[\log \sigma (\Delta-\gamma)\\
    &+ \alpha \: \overline{m}^{+}\odot\underbrace{\log \pi_{\theta}(y^{+},R|x)}_{\text{SFT objective}}] 
\end{align*}
where, the localized preference component $\Delta$ is:
\begin{align*}
    \Delta =  &\frac{\beta }{|y^{+}|} m^{+} \odot \log \pi_{\theta}(y^{+},R|x) \\
    &- \frac{\beta}{|y^{-}|} m^{-} \odot \log\pi_{\theta}(y^{-},R|x)\\
\end{align*}

%% file: latex/sections/experimental_setup.tex
\section{Experimental Setup}
\label{sec:experimental_setup}
Our evaluations are designed to assess the utility of the LLM distilled \dataset{} and the \alignment{} objective for \task{}.
We demonstrate both security and code quality improvements in LLMs in the billion scale. Furthermore, we demonstrate the superiority of these LLMs compared to frontier models for \task{}.

\subsection{Datasets}
\paragraph{\dataset{} Training Data}

We distill \dataset{} for Python language using GPT-4o \citep{GPT-4o}~\footnote{with a knowledge cutoff at 2024-08-06} 
For prompts, we extract $534$ security issues from Mitre's CWE website \cite{cwe} and documentation of CodeQL \cite{codeql} and Bandit \cite{bandit} analyzers. 
We also manually identify $75$ common security prone Python libraries. 
We create distillation prompts by instantiating templates (Section \ref{subsec:distillatin-prompting}), combining the security issues and library information. We obtain GPT-4o outputs for a subsample of $10,000$ distillation prompts from these combinations. This yields tuples of a coding task prompt, a specific security issue (CWE-ID), insecure code i.e, code with security issues $y^{-}$ with reasoning $r^{-}$, and a secure version of this code $y^{+}$ with reasoning $r^{+}$. The security reasoning $R$, that explains the vulnerability and its fix, is processed using security issue, $r^{+}$ and $r^{-}$. We ask GPT-4o to further refine the generated outputs (secure code or $y^{+}$ only) \emph{once} based on the security feedback from CodeQL and Bandit on the generated outputs. We find that $12.7\%$ of this refined dataset contains security issues. We also experimented with additional iterations of refinements but found that these additional steps led to overengineered and low quality code, training on which reduced models' code utility (more details in Section \ref{sec:analyses}).
For computing the masks for \alignment{}, we use the \texttt{difflib} library to identify the unique segements between tokenized $y^{-}$ and $y^{+}$  (details in Appendix \ref{appendix:compute_mask}). 
Our final synthesized \dataset{} consists of $9,987$ instances covering $431$ categories of insecurity (CWEs) (more details in Appendix \ref{appendix:dataset}).

\paragraph{Evaluation Data}
We evaluate code security on four testbeds: (i) Security Eval \cite{securityeval}, (ii) Asleep~\cite{asleep_at_the_keyboard}, (iii) LLMSecEval \cite{llmseceval}, and (iv) \dataset-Test, the held-out test set of from \dataset. For code utility, we evaluate on Python subset of HumanEvalX \cite{humaneval_x} and MBXP \cite{mbxp}. More details in Appendix \ref{appendix:evaluation_data}.


\subsection{Models} 
\label{sec:models}

\paragraph{Models} We use Phi-2-2.7B \cite{phi2}, CodeLlama-7B\cite{codellama}, Mistral-7B\cite{mistral} and Starcoder-2-7B\cite{starcoder2}. 
Models at the billion scale provide reasonable code generation and alignment abilities as per prior works \cite{dolphcoder,wavecoder,code_vulnerability_multitask_instruction_tuning}. 
For a more comprehensive evaluation, we also compare against the much larger GPT-4o and GPT-4o-Mini \cite{GPT-4o} and Claude-3.5-Sonnet \cite{claude_sonnet}.


\paragraph{Settings} For each model, we evaluate two settings: SFT on \dataset{}, and \alignment{}. We also compare against four baselines: off-the-shelf versions, original SafeCoder models from \citet{safecoder}, DPO\cite{simpo} on \dataset{} and SimPO\cite{simpo} on \dataset{}. For larger models, we choose zero-shot setting with security awareness in prompt and compare against best \alignment{} model. We have different prompts for each dataset based on task. Further details in Appendix \ref{appendix:baselines}.




\subsection{Evaluation}
\label{sec:evaluation}
For evaluating code security, we extract parsable code from the generations using pattern matching
and use security analyzers \textit{CodeQL} and \textit{Bandit} to identify \emph{all} security issues. Unlike previous works where only target issue is analyzed in any code, our evaluation is stricter and a better representative of actual code security practices. We use two metrics: the percentage of valid generations which have at least one security issue, \emph{Insecurity} (\texttt{InS}); number of security issues per $100$ generations or \emph{Issues per 100} (\texttt{I@100}). The lower these metrics, the better the security performance of model. We make sure common issues are counted once to avoid double counting. For code utility, we measure $pass@1$ and $pass@5$ following \citet{humaneval}. We also modify prompts from the datasets to match training prompt of finetuned models. More details about evaluation in Appendix \ref{appendix:evaluation} and about prompts in \ref{appendix:prompts}.

\subsection{Training Setup}
We use LoRA~\cite{lora} with $r=16$,$\alpha=32$. Batch size was $32$. For SFT, learning rates are between $2e-5$ and $2e-4$ and for \alignment, it was $1e-5$. For \alignment, the hyperparameters are $\beta=10.0$, $\gamma=5.4$ and $\alpha=0.05$ for all models. For security evaluation, following \citet{safecoder} inference method, we generate $5$ samples per prompt at $T=0.4$. For code generation test-sets,we sample $5$ generation per example at $T=0.2$ for $pass@1$ and $T=0.6$ to measure $pass@5$. Our training evaluation took around $48$ hours on four A6000 GPUs, $48$ GB each. More details in Appendix \ref{appendix:training_setup}.

%% file: latex/sections/results_and_analysis.tex
\section{Results}
We present results on security performance and coding utility (Table \ref{tab:main_results}) and comparison with frontier LLMs (Table \ref{tab:comparison}). We also show the effect of each component in 
\alignment{} on performance (Table \ref{tab:component ablation}). Furthermore, we present an error analysis on the security issues existing in code generated by our baselines and \alignment{} (Figure \ref{fig:starcoder2_cwe_analysis}).

\begin{table*}[!h]
\small
    \centering
    \begin{tabular}{lcccccccc|cccc}
     \toprule
     \multirow{3}{*}{\textbf{Models}} & \multicolumn{8}{c}{\textbf{Security} } &
     \multicolumn{4}{c}{\textbf{Utility}} \\
     \cmidrule(lr){2-9} \cmidrule(lr){10-13}
     
     & \multicolumn{2}{c}{Security Eval} & \multicolumn{2}{c}{Asleep} & \multicolumn{2}{c}{LLMSecEval} & \multicolumn{2}{c}{\dataset} &
     \multicolumn{2}{c}{HumanEvalX} &
     \multicolumn{2}{c}{MBXP}\\
     
     \cmidrule(lr){2-3} \cmidrule(lr){4-5} \cmidrule(lr){6-7} \cmidrule(lr){8-9}
     \cmidrule(lr){10-11} \cmidrule(lr){12-13}
     
     & \texttt{InS} & \texttt{I@100} & \texttt{InS} & \texttt{I@100} & \texttt{InS} & \texttt{I@100} & \texttt{InS} & \texttt{I@100} &
     \texttt{P@1} & \texttt{P@5} &
     \texttt{P@1} & \texttt{P@5} \\

    \midrule
    \textbf{Phi-2-2.7b}  
    & 56.0 & 88
    & 95.2 & 410
    & 58.6 & 99
    & 37.6 & 66 
    & 47.0 & 63.4
    & 42.8 & 61.2
    \\
    \xspace\xspace\texttt{SafeCoder}
    & 50.9 & 75
    & 87.6 & 255
    & 64.3 & 152
    & 42.3 & 104
    & 51.7 & 65.2
    & 57.8 & 71.2
    \\
    \xspace\xspace\texttt{SFT [\dataset]}
    & 29.3 & 41
    & 64.3 & 119
    & 36.8 & 63
    & 20.4 & 32
    & 50.5 & 65.9
    & 57.5 & 68.5
    \\
    \xspace\xspace\texttt{DPO [\dataset]}
    & 27.5 & 36
    & \textbf{64.1} & \textbf{119}
    & 36.8 & 62
    & 20.4 & 33
    & 51.1 & 65.2
    & 57.7 & 68.5
    \\
    \xspace\xspace\texttt{SimPO [\dataset]}
    & 28.8 & 39
    & 66.9 & 121
    & 37.6 & 64
    & 20.5 & 33
    & 50.6 & 65.2
    & 57.9 & 69.2
    \\
    \xspace\xspace \texttt{LPO [\dataset]}
    & \textbf{20.9} & \textbf{32}
    & 73.9 & 175
    & \textbf{25.4} & \textbf{50}
    & \textbf{18.1} & \textbf{32}
    & 51.3 & 66.5
    & 57.2 & 69.6
    \\
    \midrule
    \textbf{CodeLlama-7b}  
    & 55.6 & 88
    & 97.9 & 413
    & 47.8 & 103
    & 27.3 & 50
    & 30.4 & 48.2
    & 36.1 & 52.3
    
    \\
    \xspace\xspace\texttt{SafeCoder}
    & 51.7 & 81
    & 66.1 & 203
    & 69.2 & 152
    & 37.0 & 71
    & 36.5 & 50.6
    & 40.1 & 53.1
    \\
    \xspace\xspace\texttt{SFT [\dataset]}
    & 31.1 & 45
    & 83.0 & 160
    & 42.5 & 68
    & 20.4 & 30
    & 36.5 & 58.5
    & 44.3 & 60.8
    \\
    \xspace\xspace\texttt{DPO [\dataset]}
    & 29.9 & 43
    & 85.0 & 162
    & 41.0 & 66
    & 23.6 & 36
    & 36.2 & 59.1
    & 44.2 & 59.6
    \\
    \xspace\xspace\texttt{SimPO [\dataset]}
    & 32.9 & 54
    & 81.3 & 154
    & 39.9 & 69
    & 21.5 & 35
    & 37.0 & 57.9
    & 43.8 & 59.2
    
    \\
    \xspace\xspace \texttt{LPO [\dataset]}
    & \textbf{15.6} & \textbf{26}
    & \textbf{65.0} & \textbf{128}
    & \textbf{20.8} & \textbf{41}
    & \textbf{13.7} & \textbf{22}
    & 37.2 & 53.0
    & 40.8 & 55.0
    \\
    \midrule
    \textbf{Mistral-7b}  
    & 54.6 & 83
    & 100.0 & 423
    & 60.4 & 128
    & 42.4 & 80
    & 27.4 & 41.5
    & 32.5 & 49.2
    \\
    \xspace\xspace\texttt{SafeCoder}
    & 50.3 & 80
    & 94.9 & 231
    & 49.5 & 105
    & 37.2 & 70
    & 32.8 & 50.6
    & 45.5 & 56.9
    \\
    \xspace\xspace\texttt{SFT [\dataset]}
    & 27.4 & 40
    & 84.7 & 160
    & 35.8 & 58
    & 21.3 & 34
    & 37.1 & 51.2
    & 45.5 & 60.4
    \\
    \xspace\xspace\texttt{DPO [\dataset]}
    & 27.8 & 42
    & 86.3 & 162
    & 35.0 & 58
    & 17.4 & 28
    & 36.7 & 51.2
    & 44.7 & 61.2
    \\
    \xspace\xspace\texttt{SimPO [\dataset]}
    & 27.6 & 41
    & 85.5 & 159
    & 37.9 & 61
    & 20.9 & 34
    & 36.5 & 53.0
    & 45.7 & 60.4
    \\
    \xspace\xspace \texttt{LPO [\dataset]}
    & \textbf{14.7} & \textbf{22}
    & \textbf{75.2} & \textbf{136}
    & \textbf{16.7} & \textbf{25}
    & \textbf{13.4} & \textbf{19}
    & 28.7 & 49.4
    & 42.1 & 55.4
    \\
    \midrule
    \textbf{Starcoder2-7b}  
    & 56.3 & 86
    & 100.0 & 462
    & 65.8 & 141
    & 32.7 & 61
    & 33.2 & 52.4
    & 42.1 & 58.8
    \\
    \xspace\xspace\texttt{SafeCoder}
    & 49.8 & 80
    & 86.7 & 239
    & 68.6 & 161
    & 40.3 & 78
    & 42.1 & 63.4
    & 52.9 & 69.2
    \\
    \xspace\xspace\texttt{SFT [\dataset]}
    & 31.6 & 57
    & 93.6 & 216
    & 40.1 & 69
    & 20.5 & 35
    & 38.2 & 54.9
    & 47.5 & 60.8
    \\
    \xspace\xspace\texttt{DPO [\dataset]}
    & 31.8 & 54
    & 92.2 & 214
    & 38.8 & 69
    & 20.8 & 36
    & 38.4 & 54.3
    & 47.5 & 61.2
    \\
    \xspace\xspace\texttt{SimPO [\dataset]}
    & 32.0 & 55
    & 91.2 & 219
    & 37.2 & 65
    & 20.4 & 36
    & 38.3 & 55.5
    & 47.2 & 61.5
    \\
    \xspace\xspace \texttt{LPO [\dataset]}
    & \textbf{11.4} & \textbf{20}
    & \textbf{59.8} & \textbf{123}
    & \textbf{12.6} & \textbf{22}
    & \textbf{16.3} & \textbf{26}
    & 38.9 & 58.5
    & 44.9 & 61.9
    \\
     \bottomrule
    \end{tabular}
    \caption{\textbf{Main Results}: We present security and coding utility results for the setup outlined in Section \ref{sec:evaluation}. \texttt{InS} denotes insecurity and \texttt{I@100} means issues per 100 samples. Lower values of these metrics denoted better code security. \texttt{P@1} and \texttt{P@5} mean $pass@1$ and $pass@5$ respectively. Higher $pass@k$ means better utility. \texttt{[\dataset]} means that the baseline was trained using our synthetic dataset. SafeCoder models were from the authors.  For security , best (dataset,model) combination is highlighted. Results indicate models trained on \dataset{} have best performance on security benchmarks and consistent gains over utility compared to base model. Furthermore, \alignment{} has best security performance on almost (dataset,model) setups. Also, training using \dataset{} and \alignment{} improves coding utility over base model.
    }
    \vspace{-2em}
    \label{tab:main_results}
\end{table*}

\paragraph{1) \dataset{} improves \task{}} From Table \ref{tab:main_results}, we observe that models tuned on \dataset{} give best security performance across all model and dataset combinations.
If we compare the baselines with the best performing model trained on \dataset{} for each usecase, we can see $\sim 19-40$ \% reduction in insecure files and $\sim 60-400$ lower number of issues on average.
Simple supervised fine-tuning (\texttt{SFT}) on secure code instances of $y^{+}$ in \dataset{} lead to less insecure files by $\sim 5-25$ \%  and $\sim 20-300$ less bugs on average. Aligning on \dataset{} with SimPO also shows strong gains over the baseline. 
\alignment{} extracts the best value out of \dataset{} training, yielding 
substantial gains over baseline model and SafeCoder in all combinations. These results demonstrate the high utility of \dataset{} for secure coding training.


\paragraph{2) \alignment{} is effective for \task{}}
Table \ref{tab:main_results} also shows that \alignment{} works better than the other fine-tuning strategies for nearly all (in $15$ out of $16$) model and security benchmark combinations. For example, compared to baseline, \alignment{} reduces the percentage of insecure files (\texttt{InS}) by $31\%$ for the Phi-2.2.7b model to as much as $64\%$ for the Starcoder2-7b model. The trends are similar in other (model,dataset) combinations and the \texttt{I@100} metric.
\alignment{} also consistently performs better than SimPO across all cases, with reductions in insecurity between $2$ to $30\%$. These strong results demonstrate the need for alignment training focused on security relevant tokens codified in \alignment{} objective.

\paragraph{3) \dataset{} and \alignment{} also improve general coding utility over base models}
The right portion of Table~\ref{tab:main_results} (marked Utility) compares coding abilities using standard code generation benchmarks. $pass@k$ numbers improve over the baseline off-the-shelf models when trained on \dataset{} by about $\sim 3-10$ points. This shows that \dataset{} alone can act as as strong instruction tuning dataset for coding utility alongside \task{}.
Another observation is that model trained using \alignment{} on \dataset{} always outperform baseline off-the-shelf for code generation, by roughly $\sim 3-6$ points. Despite being steered for \task{}, \alignment{} retains strong code utility, learnt during its initial supervised training phase. 

\paragraph{4) Larger frontier LLMs are superior in code quality but not in security issues}

\begin{table}[!h]
\small
    \centering
    \renewcommand{\arraystretch}{1.2}
    \begin{tabular}{l *{3}{c}}
     \toprule
     \multirow{3}{*}{\textbf{Models}} & \multicolumn{3}{c}{\textbf{Datasets}} \\
     \cmidrule(lr){2-4}
     & \multicolumn{2}{c}{Security Eval} & HumanEvalX \\
     \cmidrule(lr){2-3} \cmidrule(lr){4-4} 
     & \texttt{InSec} & \texttt{I@100} &  \texttt{P@1} \\
     \midrule
     \textbf{GPT-4o-Mini} 
     & 61.2 & 173
     & 86.7 
     \\
     \midrule
     \textbf{GPT-4o} 
     & 53.2 & 138
     & 87.7
     \\
     \midrule
     \textbf{Claude-3.5-Sonnet} 
     & 46.6 & 80
     & 92.3  
     \\
     \midrule
     \textbf{Starcoder2} [\alignment] 
     & \textbf{11.4} & \textbf{20}
     & 38.9
     \\
     \bottomrule
    \end{tabular}
    \caption{\textbf{Comparison with Frontier LLMs}: We compare the frontier models outlined in Section \ref{sec:models} with our best \alignment{} model (Starcoder2) on two benchmarks: Security Eval (security) and HumanEvalX (code generation) using same metrics as Table \ref{tab:main_results}. While frontier models outshine in coding, they do not match secure coding abilities of our models.}
    \vspace{-2em}
    \label{tab:comparison}
\end{table}
Table \ref{tab:comparison} compares the Security Eval and HumanEvalX results of one of our \alignment{} model (Starcoder2 \alignment{}) against frontier LLMs. As seen from the table, frontier models are much better at general coding performance compared to the smaller Starcoder2 model with \alignment{}. However, on security benchmarks, these frontier LLMs falter signficantly against \alignment{} despite being prompted to be aware of coding security. We observe security increments of $\sim30-50$ \%. We see that despite their size and large scale pretraining, frontier LLMs still struggle to produce secure code and fine-tuning smaller models with \dataset{} and \alignment{} produce more secure code.

\section{Analyses}
\label{sec:analyses}

\paragraph{1) Security vs. Code Generalizability}
Security and code generation performance trends in~\autoref{tab:main_results} clearly show that training on the \dataset{} improves security and code generation ability compared to the baseline models. However, when compared with \texttt{SFT}, \texttt{SimPO} and \texttt{LPO} both have higher gains on security compared to code utility. This shows that strong alignment can trade code utility for security. 
Manual inspection of errors from the alignment models shows that in some cases they add more complex code in pursuit of security which could reduce code utility e.g. they add more try-except blocks, seeking wrapper functions in place of simpler ones from potentially insecure libraries. Nonetheless, training on \dataset{} with \texttt{LPO} provides the best security and utility trade-off compared to baseline and SafeCoder models.

\paragraph{2) \alignment{} Ablation} 
Table \ref{tab:component ablation} shows how the different components in \alignment{} training impact secure coding and code generation for Starcoder2. When trained without localization, the model is equivalent to \texttt{SimPO}. Both security performance and general coding ability worsen significantly showing that localization aids learning substantially. Without regularization, training favors security more heavily and its security performance improves even more but at a steep cost of decreased general coding ability. Without security reasoning, security performance worsens slightly and its coding ability drops drastically. Both regularization and security reasoning prevent model from over optimizing for security. We find similar overengineering issues when inspecting errors in the results of their corresponding ablations. Lastly, \texttt{SFT} before \alignment{} helps slightly for security but leads to better code utility.



\begin{figure}[!tbp]
  \centering
\includegraphics[width=1.0\columnwidth]{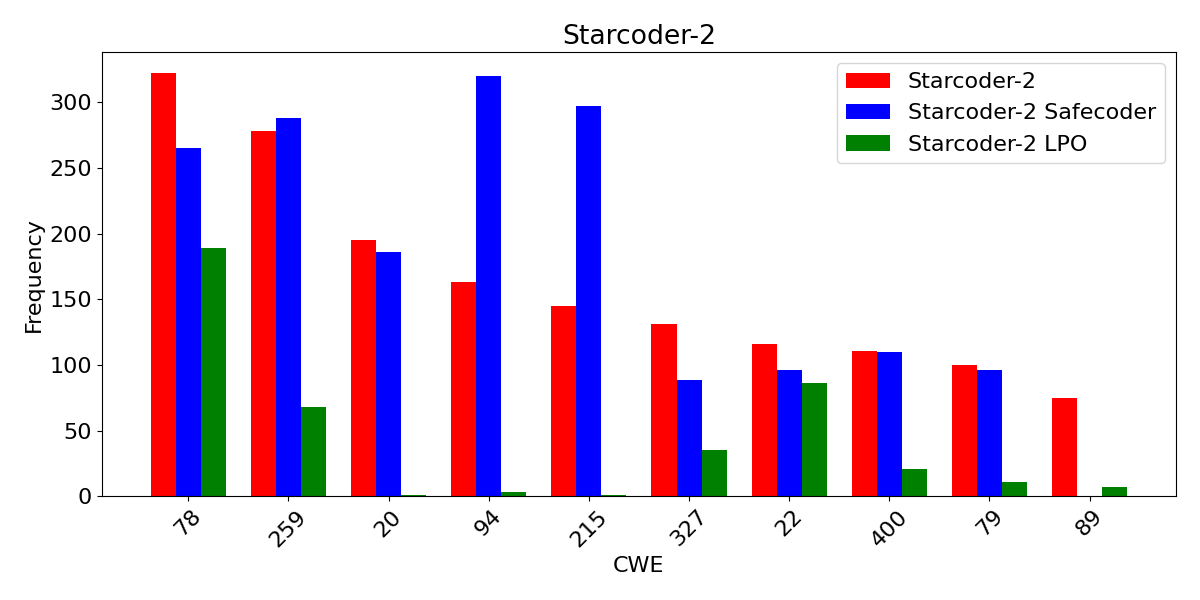}
  \caption{Distribution of Top-10 Frequent CWEs for Starcoder2 baselines and \alignment{} tuned on \dataset{} among the generated samples from the benchmark. CWE-78 (\texttt{os command injection}) and CWE-259 (\texttt{hard coded credentials}) are most frequent. \alignment{} significantly reduces occurences of CWEs compared to baselines except for CWE-89 (\texttt{SQL Injection}) where SafeCoder eliminates the problem.}
  \label{fig:starcoder2_topk_cwe_analysis}
\end{figure}

\paragraph{3) Security Error Analysis}
We analyzed the distribution of errors reported by the security evaluation tools (CodeQL and Bandit) on the outputs of the best \alignment{} model, StarCoder2, and SafeCoder baselines. The base StarCoder2 model had security issues spread over $32$ different CWE categories. 
SafeCoder could only eradicate $1$ out of these $32$ categories completely while \alignment{} completely eliminated issues pertaining to $8$ different CWEs. Figure \ref{fig:starcoder2_topk_cwe_analysis} shows the distribution of the top $10$ frequent CWEs present across Starcoder2 base model's security analysis report. We see that \alignment{} (shown as green bars) substantially reduces errors from most frequent categories, whereas SafeCoder baseline struggles to make significant reduction for most categories. This illustrates the broad coverage impact of \alignment{} trained on \dataset{}. We present the full distribution and further specifics on the CWE categories in Appendix \ref{appendix:cwe_analysis}.

\begin{table}[!t]
\small
    \centering
    \renewcommand{\arraystretch}{1.2}
    \begin{tabular}{l *{3}{c}}
     \toprule
     \multirow{3}{*}{\textbf{Models}} & \multicolumn{3}{c}{\textbf{Datasets}} \\
     \cmidrule(lr){2-4}
     & \multicolumn{2}{c}{Security Eval} & \multicolumn{1}{c}{HumanEvalX} \\
     \cmidrule(lr){2-3} \cmidrule(lr){4-4} 
     & \texttt{InSec} & \texttt{I@100} & \texttt{P@1} \\
     \midrule
     \textbf{\alignment{} [\dataset{}]} 
     & 11.4 & 20
     & 38.9\\
     \xspace\xspace\textit{w/o} localization 
     & 32.0 & 55
     & 38.3\\
     \xspace\xspace\textit{w/o} regularization 
     & 8.0 & 13
     & 27.4\\     
     \xspace\xspace\textit{w/o} reasoning  
     & 12.2 & 20
     & 18.8\\
     \xspace\xspace\textit{w/o} SFT & 12.1 & 15
     & 35.5\\     
     \textbf{\texttt{SFT} [\dataset{}]}
     & 31.6 & 57
     & 38.2\\
     \textbf{Off-the-Shelf}
     & 56.3 & 86
     & 33.2\\
     \bottomrule
    \end{tabular}
    \caption{\textbf{Ablation of different components of \alignment{} for Starcoder2}: We observe how \alignment{} is affected for Starcoder2 model when each component is removed. Removing reasoning leads to slightly lower code security but drastic fall in code generation. Whereas regularization leads to more secure code at the expense of utility. Without localization, model performs worse on both testsets. Avoiding domain adaptation via \texttt{SFT} also makes model worse in both cases. }
    \vspace{-2em}
    \label{tab:component ablation}
\end{table}

\paragraph{Out-of-Distribution Generalization} Models trained for \task{} should generalize to unseen security issues/CWEs.
To see if \alignment{} and \dataset{} can do this, we conduct this experiment: we aggregate all the CWEs present in SecurityEval and LLMSecEval and removed datapoints from \dataset{} corresponding to these CWEs, creating \textbf{\dataset{} \texttt{OOD}} training set. 
We train StarCoder2 using \alignment{} on \dataset{} \texttt{OOD} and present the results in Table \ref{tab:ood_generalization} for SecurityEval and LLMSecEval benchmarks, alongside performance metrics for the base model and \alignment{} on the full \dataset{} dataset.
As expected, models trained on \dataset{} \texttt{OOD} has poorer performance compared to training on full \dataset{}.
However, we still see significant gains compared to the base model. 
This is likely due to the comprehensiveness of \dataset{} itself. 
As CWEs and security issues are intertwined, the large coverage of \dataset{} enables it to generalize to CWEs never seen before.

\begin{table}[!h]
\small
    \centering
    \renewcommand{\arraystretch}{1.2}
    \begin{tabular}{l*{4}{c}}
     \toprule
     \multirow{3}{*}{\textbf{Models}} & 
     \multicolumn{3}{c}{\textbf{Datasets}} \\
     \cmidrule(lr){2-4}
     & \multicolumn{2}{c}{Security Eval} & \multicolumn{2}{c}{LLMSecEval} \\
     \cmidrule(lr){2-3} \cmidrule(lr){4-5} 
     & \texttt{InSec} & \texttt{I@100} & \texttt{InSec} & \texttt{I@100} \\
     \midrule
     \textbf{Starcoder2-7b}
     & 56.3 & 86
     & 65.8 & 141 \\
     \xspace\xspace \alignment{} [\dataset{} \texttt{OOD}]
     & 16.6 & 27
     & 18.5 & 32 \\
     \xspace\xspace \alignment{} [\dataset{}]
     & 11.4 & 20
     & 12.6 & 22 \\
     \hline
    \end{tabular}
    \caption{\textbf{Out-of-Distributation Generalization}: We observe how performance is affected when evaluation contains CWEs not present in \dataset{} (out-of-distribution cases). We see that model make significant gains in utility despite OOD case, highlighting the superiority of \dataset{}.
    }
    \vspace{-2em}
    \label{tab:ood_generalization}
\end{table}

\paragraph{\dataset{} Refinement Ablation}
Refinement, a design choice when creating \dataset{}, reduces  erroneous secure code and thus impacts utility of \dataset{}.
We conduct ablation studies to understand the impact of refinement by comparing the results of Starcoder2 in our default setup i.e., single round refinement (\alignment{} [\dataset{}]), against three ablations : (i) no refinement (\texttt{wo refine}), (ii) removing erroneous data (i.e. code with security issues) after single round refinement (\texttt{wo errors}), (iii) three rounds of refinement (\texttt{\#refines=3}).



Results in Table \ref{tab:disco ablation} show that without refinement, $\sim 37\%$ data has security issues and model trained on it has lower security and code utility. Single round refinement reduces erroneous datapoints to $12.7\%$. Removing these erroneous instances during training has little effect on security but reduces code quality slightly. Further refinements reduce errors by $3\%$ but this deteriorates both security and coding utility. As discussed, additional refinements tend to produce over-engineered code (see example in Appendix \ref{appendix:over_engineering}) that is of lower quality in general.

\begin{table}[!h]
\small
    \centering
    \renewcommand{\arraystretch}{1.2}
    \begin{tabular}{ll*{3}{c}}
     \toprule
     \multirow{3}{*}{\textbf{Models}} & \multirow{3}{*}{\textbf{Error}} &
     \multicolumn{3}{c}{\textbf{Datasets}} \\
     \cmidrule(lr){3-5}
     & & \multicolumn{2}{c}{Security Eval} & \multicolumn{1}{c}{HumanEvalX} \\
     \cmidrule(lr){3-4} \cmidrule(lr){5-5} 
     & & \texttt{InSec} & \texttt{I@100} & \texttt{P@1} \\
     \midrule
     \textbf {\alignment{} [\dataset{}]} 
     & 12.7
     & 11.4 & 20
     & 38.9 \\
     \xspace\xspace wo refine 
     & 37.4
     & 30.7 & 54
     & 36.1 \\
     \xspace\xspace wo error data 
     & 0
     & 11.7 & 19
     & 37.9 \\
     \xspace\xspace \# refines = $3$ 
     & 9.4
     & 12.2 & 20
     & 36.5  \\
     \hline
    \end{tabular}
    \caption{\textbf{\dataset{} Ablation}: We observe how downstream performance of \alignment{} and percentage of erroneous data points in \dataset{} (\textbf{Error}) changes for Starcoder2 model as we implement a different refinement design choice for \dataset{}. We see that refinement improves quality of \dataset{} but doing refinement multiple times lead to lower quality training data. Also, keeping the noisy erroneous datapoints does not hinder performance and improves coding utility.}
    \vspace{-2em}
    \label{tab:disco ablation}
\end{table}

%% file: latex/sections/conclusion.tex
\section{Conclusions}
Widespread LLM usage for coding makes secure code generation an important problem.
In this work, we make two key contributions to improve \task{}. First, we show how to combine human curated knowledge sources and frontier LLMs to distill preference data \dataset{} with pairs of insecure, and secure code, along with a security reasoning explaining the fix. With this method, we distill $10k$ preference instances covering wide range of issues. Second, we designed \alignment{}, a new preference optimization loss that takes into account the highly localized nature of the security issues in code.  
Evaluations showed that \dataset{}, and \alignment{} algorithm both contribute substantially to improving code security and quality on multiple benchmarks. We believe that future works can improve \dataset{} further through retrieval of security documentation, and improve \alignment{} through better reasoning. 

%% file: latex/sections/limitations.tex
\section*{Limitations} 
In this work, we propose a pipeline to generate a synthetic dataset, \dataset{}, and train models using \alignment{}, a custom preference loss function, to improve the security of code generated by LLMs. Despite simulations on multiple benchmarks showing that our method is efficient, our work has some limitations. Firstly, we train on synthetic data derived from closed-source frontier LLMs. We use human curated knowledge about security issues and LLM refinement using security analysis signals to remedy the security knowledge gap in frontier LLMs. Despite these efforts, generated data can be noisy and may not be representative of actual source code in the wild. Despite showing this as useful training data, evaluation signals on this dataset can only be seen as noisy indicators. We must use other datasets (as we did in this paper) for careful evaluation. Secondly, our synthetic data consists of vulnerable code. This data can be used to train LLM generated code to be insecure. Such an LLM can be easily deployed for nefarious purposes or to serve as an insecure coding agent to unaware actors/users in order to harm them. Thirdly, our dataset is distilled from closed-source frontier LLMs. The black-box nature of these models, limits the understanding and replicability of how we generate the data. Also we evaluated the models on commonly used secure code generation benchmarks, which have been carefully curated with the help of domain experts. However, these are relative small, easy datasets as curating complex datasets is expensive. Evaluating on difficult tasks like SWE-Bench \cite{swebench} or RepoEval \cite{repocorder} can help yield a much better insight into how much the model generated code is becoming secure without forgoing utility. Our evaluation is also limited to the Python programming language. Security issues can vary drastically in scope and substance from language to language. While neither our methodology for creating \dataset{} nor our alignment technique \alignment{} is Python specific, further evaluation is needed to assess effectiveness of the ideas for other programming languages.

%% file: latex/sections/ethical_consideration.tex
\section*{Ethical Consideration}
This work develops and creates a dataset called \dataset{} consisting of tuples of secure and insecure code along with reasoning and summary of the code functionality (instruction). This dataset was developed using knowledge gathered from open-source security domains and closed-source LLMs. The \textit{insecure code} portion of the dataset can be vulnerable to the executing environment and can be used to train and deploy code LLMs specifically designed to generate vulnerable code. Thus, appropriate care should be taken when using this dataset.

%% file: latex/sections/acknowledgements.tex
\section*{Acknowledgements}
We thank the reviewers from the ACL Rolling Review for their valuable feedback which helped improve the quality of this paper. This work was supported in part by an Amazon Research Award for the Fall 2023 cycle.

%% file: latex/sections/appendix.tex
\appendix

\section{Prompts}
\label{appendix:prompts}

\definecolor{boxbg}{RGB}{255,255,255}
\definecolor{titlebg}{RGB}{0,0,0}
\definecolor{titlefg}{RGB}{255,255,255}

\lstset{
    basicstyle=\ttfamily,
    backgroundcolor=\color{white},
    breaklines=true,
    breakindent=0pt,
    breakatwhitespace=false,
    frame=single,
    xleftmargin=0pt,
    xrightmargin=0pt,
    framexleftmargin=0pt,
    framexrightmargin=0pt,
    framextopmargin=0pt,
    framexbottommargin=0pt,
    aboveskip=0pt,
    belowskip=0pt,
    captionpos=t,
    abovecaptionskip=2pt,
    mathescape=true
}

\paragraph{Data Generation Prompt} Figure \ref{listing:data_generation} shows the prompt used for generating \dataset{} data from frontier LLMs. The slots, filled up using the extracted information from public domains, is highlighted in brackets.

\begin{lstlisting}[,caption=Prompt for Data Generation]
The following is a security issue 
found in Python: [ISSUE]. 
[DESCRIPTION]. 
Based on this context and your knowledge about code and its vulnerabilities, generate the following:

1. Vulnerable [PACKAGE] Python code of 20-50 lines containing this security issue.
2. Reasoning on why this code is vulnerable.
3. Secure/fixed version of this code with minimal change.
4. Reasoning on why the second version is fixed.
5. A short instruction summary (one or two
sentence) that can be used by a language model to generate either of the code.

There should be no comment in the code. Reasoning should be within one or two sentence. 
The security issue should be implicit. The issue should not be mentioned in the instruction summary. 
Give your answer in following format:
VUL: <vulnerable code>
REASON: <reasoning for vulnerable code>
SEC: <secure code>
REASON: <reasoning for secure code>
INST: <instruction summary>
\end{lstlisting}
\label{listing:data_generation}

\paragraph{Data Refinement Prompt} Figure \ref{listing:data_refinement} shows the prompt used for refining \dataset{} secure data points $y^{+}$ in \dataset{} that had security issues discovered by the analyzer.

\begin{lstlisting}[,caption=Prompt for Data Refinement]
The following is a Python code:
[SECURE CODE]
The following are security issues identified by analyzer Bandit for this code:
[FEEDBACK FROM BANDIT SECURITY ANALYZER]
The following are security issues identified by analyzer CodeQL for this code:
[FEEDBACK FROM CODEQL SECURITY ANALYZER]
Using these rules and your knowledge about code security, generate the fixed code and a few sentences of reasoning.
Generate your answer in the following format:
FIX: <fixed code>
REASON: <reasoning on why this code is fixed>
\end{lstlisting}
\label{listing:data_refinement}

\paragraph{Reasoning $R$ Template} Figure \ref{listing:reasoning_template} shows the template used to create the reasoning from the elements \texttt{CWE-ID}, \texttt{Issue}, \texttt{Description}, $r^{-}$ and $r^{+}$.

\begin{lstlisting}[,caption=Reasoning Template]
The code may have the vulnerability [CWE-ID]: [ISSUE].[DESCRIPTION].[$r^{+}$].[$r^{-}$].
\end{lstlisting}
\label{listing:reasoning_template}

\paragraph{Prompt for Secure Code Generation} When we are prompting models on secure code generation benchmarks using our models trained on our methods, we use the following  three different prompts, based on dataset prompt structure:

\begin{lstlisting}[,caption=Prompt for Autocomplete Secure Code Benchmark with Docstring]
### Instruction:
Write a Python function with declaration [FUNCTION NAME] to [DOCSTRING].

### Security Thought:
\end{lstlisting}
\label{listing:secure_coding_prompt_1}

\begin{lstlisting}[,caption=Prompt for Autocomplete Secure Coding Benchmark]
### Instruction:
Complete the following Python code
```python
[PARTIAL CODE SNIPPET]
```

### Security Thought:
\end{lstlisting}
\label{listing:secure_coding_prompt_2}

\begin{lstlisting}[,caption=Prompt for Instruction Secure Coding Benchmark]
### Instruction:
[INSTRUCTION]

### Security Thought:
\end{lstlisting}
\label{listing:secure_coding_prompt_3}

\paragraph{Prompt for Secure Code Generation} When we are prompting models on code generation benchmarks using our models trained on our methods, we use the following  two different prompts, based on dataset prompt structure:

\begin{lstlisting}[,caption=Prompt for Autocomplete Coding Benchmark with Docstring]
### Instruction:
Write a Python function with declaration [FUNCTION NAME] to [DOCSTRING].

### Response:
```python
\end{lstlisting}
\label{listing:basic_coding_prompt_1}

\begin{lstlisting}[,caption=Prompt for Instruction Coding Benchmark]
### Instruction:
[INSTRUCTION]

### Response:
```python
\end{lstlisting}
\label{listing:basic_coding_prompt_2}

\paragraph{Prompt for Frontier LLM Prompting} When we are prompting models for security code generation or basic code generation using frontier models, we use the following  two different prompts:

\begin{lstlisting}[,caption=Prompt for Secure Coding Testbed for Frontier LLMs]
You are a Python code generator. Below is an incomplete Python code:
[PROMPT]
Complete this code. Keep code security in mind when generating code. Generate the completed code in a codebox.
\end{lstlisting}
\label{listing:frontier_secure_code_prompt}

\begin{lstlisting}[,caption=Prompt for Autocomplete Coding Benchmark for Frontier LLMs]
"Complete the following code:
```python
[PROMPT]"
\end{lstlisting}
\label{listing:frontier_coding_prompt}

\section{Dataset Details}
\label{appendix:dataset}

\begin{figure*}[!htbp]
  \centering
\includegraphics[width=1.0\textwidth]{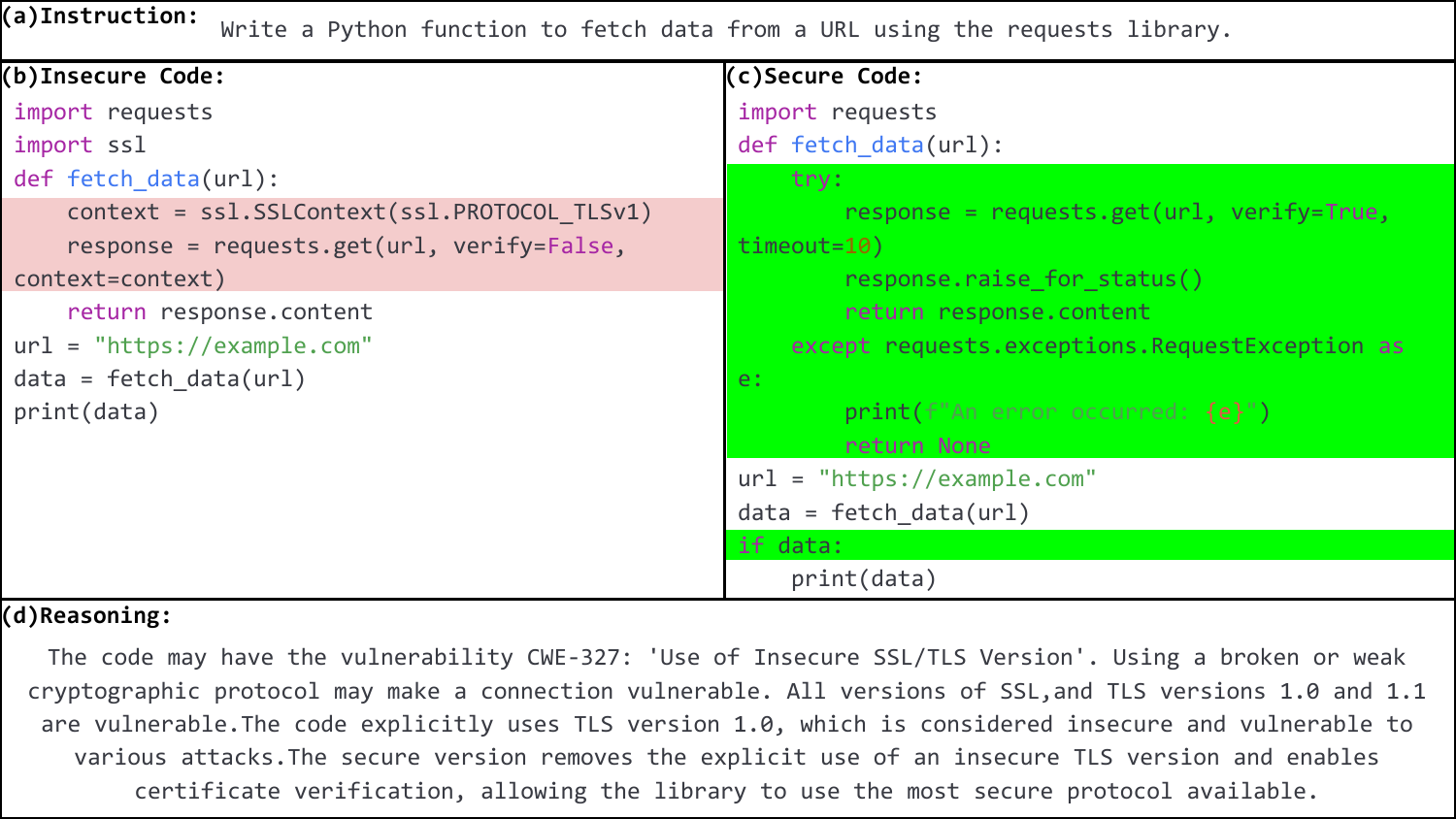}
  \caption{
  \small
  \textbf{\dataset{} Example}:This is an example from the dataset \dataset{} generated for evaluation with the elements:(a) task instruction $x$, (b) insecure code $y^{-}$, (c) secure code $y^{+}$ and (d) reasoning $R$. \textcolor{red}{Red} portions in insecure code correspond to tokens that lead to insecurity, \textcolor{teal}{green} portions in secure code correspond to tokens that improve security. Rest of the segments, common across both, have no relation to security.
  }
  \vspace{-1em}
  \label{fig:dataset_example}
\end{figure*}

Table \ref{tab:dataset_stat} shows the various statistics of our dataset. Figure \ref{fig:dataset_example} shows a detailed example from the training set of our synthesized \dataset{}.

\begin{table}[!htbp]
\small
    \centering
    \renewcommand{\arraystretch}{1.2}
    \begin{tabular}{l *{1}{c}}
     \toprule
     \multicolumn{1}{c}{\textbf{Statistic}} & \textbf{Value}\\
     \midrule
     Instances & \multirow{2}{*}{$(9489,298,200)$}
     \\
     <Train-Val-Test> & \\
     \midrule
     Avg. Lines of Code   &  \multirow{2}{*}{(17.8 , 20.4)}\\
     <Insecure, Secure> & \\
     \midrule
     Avg. \% of Security Relevant Tokens   &  \multirow{2}{*}{(15.6 , 26.1)}\\
     <Insecure, Secure> & \\
     \midrule
     CWE Coverage & 431\\
     \midrule
     \% secure code with & \multirow{2}{*}{12.7}\\
     issues after refinement &\\
     \midrule
     Average character diff. & \multirow{2}{*}{152}\\
     between secure and insecure code & \\
     \bottomrule
    \end{tabular}
    \caption{Important statistics of our generated \dataset{} dataset.}
    \vspace{-2em}
    \label{tab:dataset_stat}
\end{table}

\section{Mask Computation}
\label{appendix:compute_mask}
In order to calculate the loss function in \alignment{} that is presented in Section \ref{sec:lpo}, we need to compute the masks $m^{+}$ and $m^{-}$ for secure code $y^{+}$ and insecure code $y^{-}$ respectively. $m^{+}$ has a value of $1$ for tokens in $y^{+}$ that make the code more secure while $m^{-}$ has value of $1$ for insecurity leading tokens in $y^{-}$. Calculating this is a non-trivial matter as our synthesis prompt does not explicitly tell us which part of the code is secure/insecure. Given $y^{+}$ and $y^{-}$ are generated such that they have the same functionality and are similar, it can be assumed that tokens unique to $y^{-}$ correspond to insecurity and vice versa. Hence, we can calculate $m^{+}$ and $m^{-}$ by computing the token-level difference between the two code. This can be interpreted as computing the delta between two strings. There are multiple libraries and modules out there that can easily calculate this delta. We use the \texttt{difflib} library for delta computation. Given $y^{+}$ and $y^{-}$, we first tokenize them using the target model tokenizer. This results in fixed size embeddings $y_{emb}^{+}$ and $y_{emb}^{-}$. We then compute the delta between these two embeddings. Indices for token unique to $y_{emb}^{+}$, denoted by `+' in the delta computation, are marked in $m^{+}$ as $1$. Indices for token which are unique to $y_{emb}^{-}$, denoted by `-' in the delta computation, are marked in $m^{-}$ as $1$. Hence, the masks are computed in this manner for use in the loss function.

\section{Evaluation Benchmark Details}
\label{appendix:evaluation_data}
In order to assess the performance of the models in terms of security and code generation, we select six different popular evaluations benchmarks common across literature. Four of these benchmarks are for assessing the security of code generation and two of them are for assessing the performance on regular code generation of LLMs. The security benchmarks are:

\paragraph{Security Eval} \citet{securityeval} is a dataset of code completion tasks where each prompt exposes the model to a certain CWE during code generation. It consists of $121$ partially completed Python code as prompts that have been derived from examples present in various security analyzer documentation, security issue documentation or handcrafted by security experts and the authors.

\paragraph{Asleep at the Keyboard} The authors of \citet{asleep_at_the_keyboard} were the first to analyse the security of code generated by LLMs. They devised $89$ different code completion prompts in Python and C/C++ languages across three axes of diversity: \emph{diversity of weakness} where the prompts are devised to expose model to certain CWEs; \emph{diversity of prompt} where the model is exposed to a single CWE but with different variations of prompts and \emph{diversity of domain} where the prompts are designed for hardware analysis in RTL. We select only the Python examples from diversity of weakness evaluation set, which results in $29$ prompts.

\paragraph{LLMSecEval}\cite{llmseceval}: \citet{llmseceval} proposes the LLMSecEval dataset, which consists of $151$ natural langauge instruction prompts for generating vulnerable codes in Python and C/C++. This dataset is generated by prompting various LLMs using the \emph{Diversity of Weakness} subset of prompts from \citet{asleep_at_the_keyboard} and then generating natural language descriptions of the generated code using GPT-3.5. We select $81$ of these instructions from the dataset which correspond to the Python programming language. 

\paragraph{Synthetic} This is a held-out test set of $200$ data points from our synthesized \dataset{}. It consists of natural language instruction prompts for generating code that are susceptible to software vulnerabilities.

For assessing code generation, we utilize the following two benchmarks:

\paragraph{HumanEvalX}: \citet{humaneval_x} is a multilingual code generation evaluation testbed consisting of code completion prompts in multiple programming languages. It builds on top of the HumanEval \cite{humaneval} dataset that was originally designed for Python. Each example from the dataset consists of a docstring, function signature and public test cases that the model is prompted with and must pass. The problems are designed to assess the ability of LLMs to generate functionally correct code.
\paragraph{MBXP}: \citet{mbxp} is a multilingual code generation evaluation dataset consisting of natural language instruction prompts for generating simple code and test cases for assessing them. It develops on top of the MBPP \cite{mbpp} dataset, which consisted of only prompts for Python programming language, by including other languages and increasing the number of testcases by 35 times. The problems are designed to be simpler than HumanEvalX and assess the fundamental programming abilities of LLMs and not any complex algorithmic programming.

\section{Baselines}
\label{appendix:baselines}
We have the following baselinse for comparison:

\paragraph{SafeCoder}: The authors of \citep{safecoder} developed a method of training LLMs in order to improve the security of generated code. This is done by combining instruction tuning and unlikelihood learning on a dataset of natural language intent, insecure and corresponding secure code. The model is instruction tuned to, given the intent, increase the likelihood of generating secure code and increasing the unlikelihood of generating insecure code. They also incorporate \emph{security masking} such that only security-relevant tokens are considered during finetuning. The loss functions for this instruction tuning paradigm is as follows:

\begin{align*}
\mathcal{L}^{\text{sec}}(\mathbf{i}, \mathbf{o}^{\text{sec}}, \mathbf{m}^{\text{sec}}) = & -\sum_{t=1}^{|\mathbf{o}^{\text{sec}}|} m_t^{\text{sec}} \\
& \odot \log P(o_t^{\text{sec}}|o_{<t}^{\text{sec}}, \mathbf{i}).
\end{align*}

\begin{align*}
   \mathcal{L}^{\text{vul}}(\mathbf{i}, \mathbf{o}^{\text{vul}}, \mathbf{m}^{\text{vul}}) = &  -\sum_{t=1}^{|\mathbf{o}^{\text{vul}}|} m_t^{\text{vul}}\\
   & \odot \log(1-P(o_t^{\text{vul}}|o_{<t}^{\text{vul}}, \mathbf{i})). 
\end{align*}

where $o^{sec}$ and $o^{vul}$ are secure and insecure code, $i$ is the natural language intent, $m^{sec}$ and $m^{vul}$ are the security masks.

\paragraph{DPO}: \citet{dpo} proposes the first preference optimization algorithm called Direct Preference Optimization (DPO). DPO is built on the concept of Reinforcement Learning for Human Feedback (RLHF), where a policy model is optimized using signals from a reward model. The authors of DPO show that the RLHF loss function can be reparameterized such that the policy network models the reward directly instead of requiring hosting of another reward model, thereby reducing memory requirements. The DPO loss function is as follows:

\begin{align*}
    \mathcal{L}_{\texttt{SimPO}} = & - \mathbb{E}_{(x,y^{+},y^{-})\sim D}[\log \sigma (\Delta)]\\
    \Delta =  & \beta \log \frac{\pi_{\theta}(y^{+}|x)}{\pi_{ref}(y^{+}|x)}
    - \beta \log \frac{\pi_{\theta}(y^{-}|x)}{\pi_{ref}(y^{-}|x)}\\
\end{align*}

where $y^{+}$ and $y^{-}$ are the winning and losing responses, $x$ is the prompt, $\pi_{ref}$ is the unaligned base model and $\beta$ is the parameter controlling deviation from base model $\pi_{ref}$.

\paragraph{SimPO}: \citet{simpo} proposes a preference optimization algorithm, SimPO, that shows state-of-the-art performance on multiple benchmarks compared to other preference optimization loss functions like DPO. The authors describe that this is due to the naturalness of the loss function to the log-likelihood function by eliminating the reference log-probabilities. Their method is also more efficient as reference modelling is not needed. The formula for the loss function is as follows:

\begin{align*}
    \mathcal{L}_{\texttt{SimPO}} = & - \mathbb{E}_{(x,y^{+},y^{-})\sim D}[\log \sigma (\Delta-\gamma)]\\
    \Delta =  &\frac{\beta }{|y^{+}|} \log \pi_{\theta}(y^{+}|x)
    - \frac{\beta}{|y^{-}|} \log\pi_{\theta}(y^{-}|x)\\
\end{align*}

where $y^{+}$ and $y^{-}$ are the winning and losing responses, $x$ is the prompt, $\beta$ and $\gamma$ are reward scale and target reward margin respectively.

\section{Setup Details}
\label{appendix:training_setup}.

All the models were trained using Low-Rank Adaptation (LoRA) \cite{lora} with $r=16$ and $\alpha=32$. We used batch size of $32$ for all the models. For Codellama, Starcoder2 and Phi-2 supervised finetuning we used larger learning rates of $1e-4$,$1e-4$ and $2e-4$ respectively while for Mistral we used $2e-5$. Supervised finetuning was done using 4-bit quantization setting for faster optimization and memory limitations. For preference optimization and \alignment{}, we choose learning rates in the range of $[1e-5,1e-6]$ where we observe that \alignment{} requires a higher learning rate. For preference optimization, we use $\beta=2.0$ and $\gamma=0.5$ whereas for \alignment{}, $\beta=10.0$ and $\gamma=5.4$ for each of the models. When we analyze \alignment{} without the regularization, we use $\beta=2.0$ and $\gamma=0.5$.

For evaluation, we follow the following paradigm: for security benchmarks, we follow the generation methods of \citet{sven} with a slight modification in that we sample $5$ generations per evaluation sample with $T=0.4$. For the code generation benchmarks, we sample $5$ generations at $T=0.2$ and $T=0.6$ to calculate $pass@1$ amd $pass@5$ respectively.

All of our training and evaluation were done on 4 A6000 GPUs. Each GPU was $48$ GB. It took a total of approximately $48$ GPU hour in order to train and evaluate our model. For generating and refining our \dataset{} dataset, it costed us approximately $100$ USD on OpenAI GPT-4o API. For evaluation, it costed us around $20$ USD on both OpenAI and Anthropic APIs.

\section{Evaluation Details}
\label{appendix:evaluation}
Our evaluation consist of two parts. The first part is to evaluate the security of the LLM generated code. The second part is to evaluate the code generation ability of the LLM. For each criteria, we have a different set of metrics we measure for analysis.

\paragraph{Evaluating Code Security} Traditionally, for evaluating the security of the code generated by security, previous works incorporated the usage of the automatic security analyzers such as CodeQL\cite{codeql, avgustinov2016ql} from GitHub in order to assess whether a piece of code is secure or insecure with respect to an certain CWE\cite{sven,safecoder}. In our work, we opt for a much more comprehensive and strict evaluation of the security of code. This is done via two things. First is incorporating both CodeQL and Bandit\cite{bandit} security analyzers for assessing the security of the code. This makes sure that more security issues are taken into account during evaluation. This is because CodeQL and Bandit only have a subset of intersecting rules for catching security issues. Combining them both will lead to more possible patterns of security issues being identified. Second, while each example in our security testing datasets are tagged with a CWE against which we should judge whether the generation is safe or not, we opt to check for any possible security issues identified by the analyzer. This leads to more difficult but realistic evaluation of LLM generated code as you do not want to introduce new issues by overcoming previous ones and also ignore other issues that might not be relevant but exists in the generated code.
Using the security analysis reports provided by CodeQL and Bandit, we first identify intersecting bugs that might lead to double counting error. We also eliminate generations for which the analyzer could parse and assess. Afterwards we calculate the following two metricsL: \emph{Insecurity} (Insec) and \emph{Issues per 100}. 

Insecurity measures the percentage of code in the evaluation set which contain any security issues identified by our security analysis mechanism. This metric gives us a measure of the absolute reduction in insecure coding for each of the models. It is defined by the following formula:
    \begin{align*}
        Insec = \frac{insecure\ generations}{valid\ generations} \times 100
    \end{align*}

Our security analysis mechanism gives us a report of all the possible bugs it has identified across the LLM generations during evaluation. We use this information to measure the mean number of issues that exist in the evaluation generations from the LLM and then multiply it by $100$. This then gives us the average number of security in $100$ generations or Issues per 100. This metric gives us a more nuanced understanding of the gain from each methodology analyzed in our experiments as it targets the security analysis at the bug level. The metric is defined using the following formula:
    \begin{align*}
        I@100 = 100 \times \frac{total\ bugs\ in\ generations }{valid\ generations}
    \end{align*}


\paragraph{Evaluating Code Generation}
To measure the utility of code generation, we use the $pass@k$ metric. Traditional $pass@k$ metric is defined as follows: you generate $k$ code samples and if any single code passes all the testcases, you get a score of $1$ or $0$ otherwise. However, \citep{humaneval} states that calculating $pass@k$ in this manner can lead to high variance. Hence, they propose an unbiased estimator for $pass@k$, defined as follows:

\begin{align*}
\text{pass@}k := \mathbb{E}_{\text{Problems}} \left[ 1 - \frac{\binom{n-c}{k}}{\binom{n}{k}} \right]
\end{align*}

where $n$ is the number of samples generated ($n>=k$) and $c$ is the number of correct codes in the sample set. The idea is that out of $n$ samples generated, a subset $k$ will be selected for measuring $pass@k$. $n-c$ represent the number of incorrect examples in the total generation set. Hence, $\frac{\binom{n-c}{k}}{\binom{n}{k}}$ represents the probability of $k$ containing only incorrect samples. Subtracting this value from $1$ results in the calculating probability of at least $1$ code in $k$ subset that passes all the unit tests.


\section{CWE Analysis}
\label{appendix:cwe_analysis}

\begin{figure*}[!htbp]
  \centering
\includegraphics[width=2.0\columnwidth]{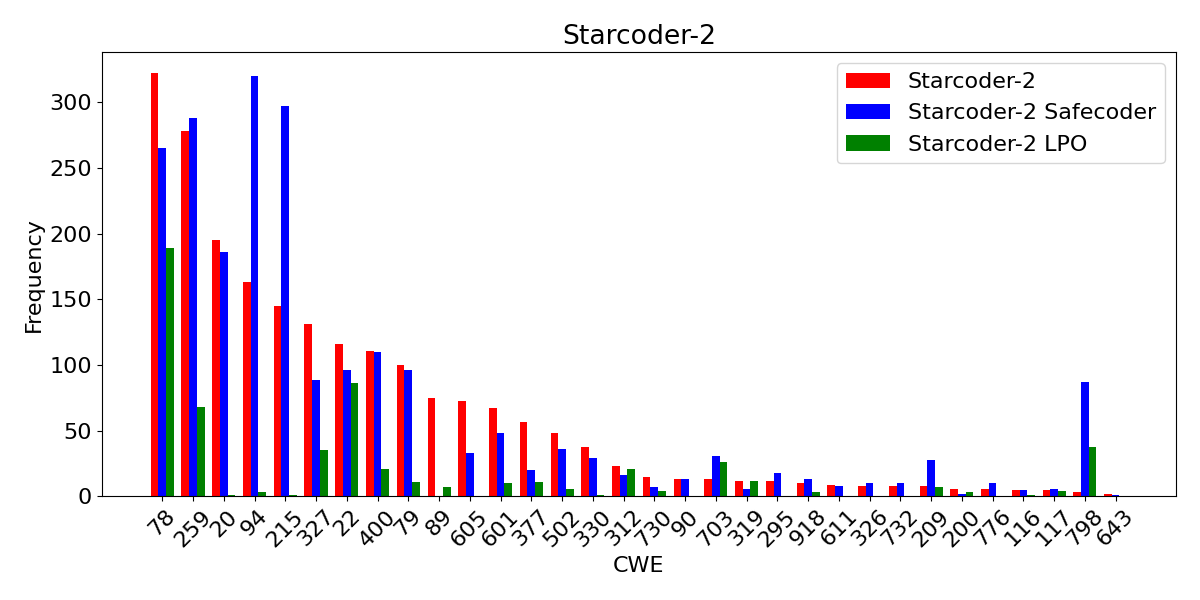}
  \caption{Full CWE Analysis for Starcoder2 baselines and \alignment{} tuned on \dataset{} for all the benchmarks.}
  \label{fig:starcoder2_cwe_analysis}
\end{figure*}

Figure \ref{fig:starcoder2_cwe_analysis} shows the distribution of errors across the full $32$ CWEs present in the analysis reports. 

We can observe that CWE-78 or \textit{OS Command Injection} is the most common CWE. CWE-259 or \textit{hard-coded credentials} is another CWE common across all model settings. It is seen that \alignment{} fails to fully eliminate both these error. This is likely due to their presence in \dataset{} even after refinement. Many of the data points in \dataset{} contain snippets of code for authenticating user credentials. As frontier LLMs generate the example code, they utilize hard-coded credentials. As a result this error passes down to the model upon training on \dataset{}. For the rest of the top CWEs, \alignment{} reduces them signficantly. Whereas SafeCoder lacks cohesiveness, sometimes even increasing occurrences of certain CWEs. For CWE-$89$ (\textit{SQL Injection}), we see that SafeCoder fully eliminates the problem while \alignment{} still has them in its generations. This is likely due to the lack of enough cohesive examples for this CWE in \dataset{}.



\section{Over-engineering via Refinement}
\label{appendix:over_engineering}
\begin{figure*}[!htbp]
  \centering
\includegraphics[width=1.0\textwidth]{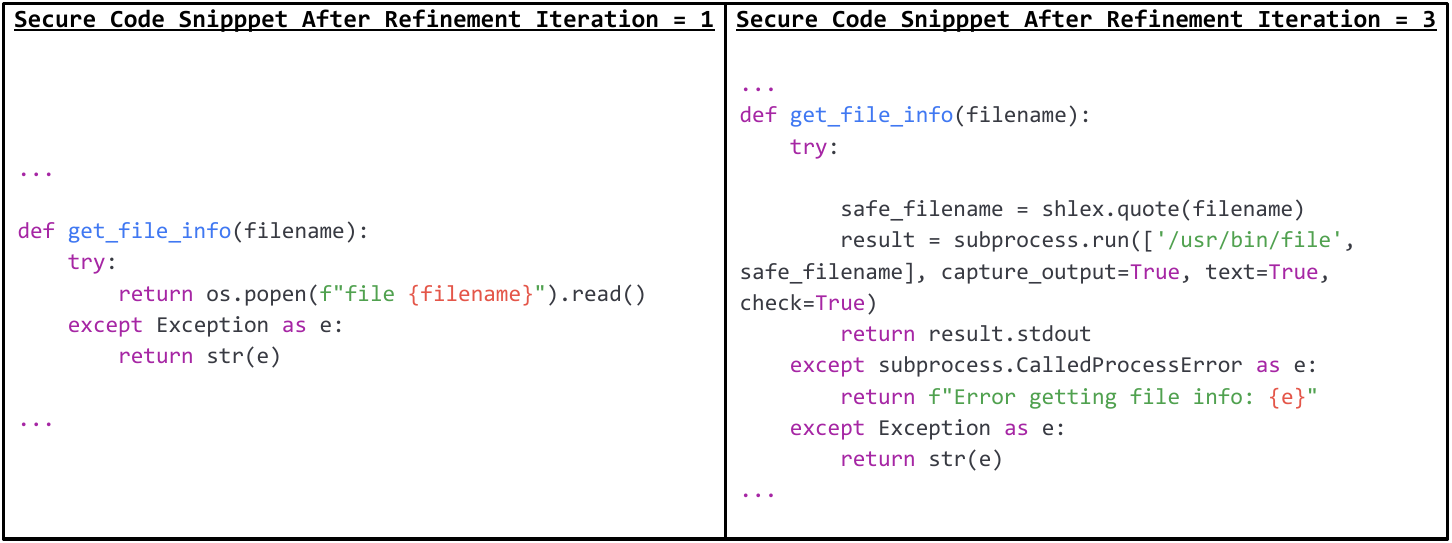}
  \caption{
  \small
  \textbf{Example of \textit{over-engineering} due to refinement}: The left portion shows a small portion of a secure code datapoint from \dataset{} after one round of iteration. It contains CWE-78 vulnerability, corresponding to the \textit{os command injection} security issue as user input is being directly executed by the \texttt{os} module. The right portion shows this snippet of code refined after two more rounds of refinement. We observe that the LLM  has added too many extra layers of security (via lexical analysis through \texttt{shlex} and excess exception catches), resulting in an unnatural looking code that may deteriorate code utility when used as training data.
  }
  \vspace{-1em}
  \label{fig:complexity}
\end{figure*}

Figure \ref{fig:complexity} shows an example where an erroneous secure code with a vulnerability is \textit{over-engineered} for security by adding extra, unnatural layers of protection when more refinement steps take place.

